\documentclass[12pt,letterpaper]{iopart}
\usepackage{graphicx}
\usepackage[T1]{fontenc}
\usepackage{iopams}

\begin{document}


\title[Exchange bias in core/shell magnetic nanoparticles]{Particle size and cooling field dependence of exchange bias in core/shell magnetic nanoparticles}

\author{\`{O}scar Iglesias, Xavier Batlle and Am\'{\i}lcar Labarta}
\address{Departament de F\'{\i}sica Fonamental and Institut de Nanociència i Nanotecnologia de la UB (IN$^2$UB), Universitat de Barcelona, Av. Diagonal 647, 08028 Barcelona, Spain}
\ead{oscar@ffn.ub.es, http://www.ffn.ub.es/oscar}

\begin{abstract}

We present a numerical simulation study of the exchange bias (EB) effect in nanoparticles with core/shell structure aimed to unveil the microscopic origin of some of the experimental 
phenomenology associated to this effect. In particular, we have focused our study on the particle size and field cooling dependence of the hysteresis loop shifts.
To this end, hysteresis loops after a field cooling process have been computed by means of Monte Carlo simulations based on a model that takes into account the peculiar properties of the core, shell and interfacial regions of the particle and the EB and coercive fields have been extracted from them. 
The results show that, as a general trend, the EB field $h_{EB}$ decreases with increasing particle size, in agreement with some experimental observations. 
However, closer inspection reveals notable oscillations of $h_{EB}$ as a function of the particle radius which we show to be closely related to the net magnetization established after field cooling at the interfacial shell spins.
For a particle with ferromagnetic interface coupling, we show that the magnitude and sign of $h_{EB}$ can be varied with the magnetic field applied during the cooling process.   
\vspace{1pc}
\end{abstract}


\vspace{2pc}
\submitto{\JPD}
\maketitle

\section{Introduction}

Magnetic nanoparticles have been the object of increasing research activity since the 1950's. Their interest range covers from permanent magnetism, magnetic recording and magnetotransport \cite{Bader_rmp06}, to macroscopic quantum phenomena or biomedical applications \cite{Tartaj_cn06}. With the appearance of new preparation techniques, magnetic particles with sizes below the 10 nm range with narrow size distributions and different compositions can be prepared thorough a variety of procedures, either embedded in a conducting or insulating matrix or diluted in organic solvents. When entering the nanometer range, magnetic nanoparticles display interesting new phenomena that can be associated either to intrinsic characteristics of the individual particle (such as finite-size and surface effects \cite{Batlle_jpd02}) or to collective properties of the nanoparticle ensamble, such as dipolar interactions \cite{Majetich_jpd06} and exchange mediated coupling between the particles. 

In the last years, particular interest has been given to particles with core/shell structure of different compositions in which usually the core is ferromagnetic (FM) and the shell is made by the parent oxide compound \cite{Nogues_physrep05}, usually an antiferromagnet (AFM), although several studies have recently reported also inverse structures \cite{Salazar_jacs07}. It is well known that proximity of a FM to an AFM induces exchange anisotropy at the interface between the two phases producing a shift of the hysteresis loop after colling in the presence of a magnetic field that has been termed exchange bias (EB). The origin of this effect has been intensively studied in thin film systems during the last decades and progress has been made in its understanding \cite{Nogues_jmmm99}. Although first reported for core/shell particles long time ago \cite{Meiklejohn_PR}, there has been a renewed interest in understanding the microscopic mechanism underlying the EB effect in these systems after the observation that the EB effect can be used to beat the superparamagnetic limit of magnetic recording media \cite{Skumryev}. Apart from the shift of the hysteresis loops, core/shell nanoparticles present peculiar magnetic phenomenology associated to EB which is not completely understood \cite{Iglesias_JNN07} due to the intrinsic inhomogeneity in the main parameters characterizing the core/shell structures present in real samples, such as particle size, shell thickness, anisotropy constants and easy axis directions. Theoretical models and theories suited for the explanation of EB in thin films bilayers are not directly applicable to core/shell nanoparticles and, in particular, are not able to account for the high values of the exchange bias fields $h_{EB}$ reported for some compositions \cite{Kiwi_jmmm01,Stamps_jpd00}. This is because the peculiar arrangement of spins at the core/shell interface of spherical or ellipsoidal nanoparticles results in intrinsic desorder and finite-size effects that are not present in the case of planar geometries.

Recently, through numerical simulations of a microscopic model of an individual nanoparticle, we have shown that the bias field $h_{EB}$ can be quantitatively related to the net local exchange fields of the interfacial shell spins that act on the particle core \cite{Iglesias_prb05,Iglesias_jpcm07}. 
Our approach has also been successful in tracing back the microscopic origin of the loop 
assymetries observed experimentally to different magnetization reversal mechanisms in both loops branches \cite{Iglesias_phab06,Iglesias_jmmm07}. 
In the present study, we focus our interest on the dependence of the EB effect on the particle size for a given value of th shell thickness and on the effect of the cooling field on the magnitude and direction of the shift of the hysteresis loops, unveiling its microscopic origin.

\section{Model and simulation method}

The Monte Carlo simulations reported here are based on the same model and methodology used previously in our recent works to which the reader is referred for the details \cite{Iglesias_prb05}.
The magnetic ions of the nanoparticle are modelled by classical spins ${\vec S}_i$ lying in the nodes of a cubic lattice and interacting through the following by the following Hamiltonian: 
\begin{eqnarray}
\label{Eq1}
{ H}/k_{B}= 
-\sum_{\langle  i,j\rangle}J_{ij}{\vec S}_i \cdot {\vec S}_j   
-k_C\sum_{i\in \mathrm{C}}(S_i^z)^2
-k_S\sum_{i\in \mathrm{Sh}}(S_i^z)^2  	
-\sum_{i= 1}^{N} \vec h\cdot{\vec S_i}
\end{eqnarray}
where in the fourth term $\vec{h}$ is the magnetic field applied along the easy-axis direction with module $\vec{h}=\mu\vec{H}/k_B$ in temperature units ($\mu$ is the magnetic moment of the spin). 
Spins in the different regions of the nanoparticle (core, shell and interface) have different magnetic properties.
In the first term, the exchange constants are FM at the core ($J_{\mathrm{C}}= 10$ K), AF at the shell ($J_{\mathrm{S}}= -5$ K) and at the interface [formed by the spins in the core (shell) having neighbours in the shell (core)], FM or AF coupling $J_{\mathrm{Int}}=\pm 0.5J_{\mathrm{C}}$ will be considered. The second and third terms correspond to the uniaxial anisotropy energy, with $k_C= 1$ and $k_S= 10$ the anisotropy constants for core and shell spins respectively.  
We will consider particles with spherical shape and different total radius $R_{\mathrm{Total}}$ ranging from $6a$ to $20a$ and constant shell thickness $R_{Sh}=3$a ($a$ is the lattice constant). 
The field cooling procedure previous to the simulation of the hysteresis loops has been performed from a temperature ($T= 20$ K) higher than the FM core ordering temperature down to $T=0.1$ in steps $\delta T= 0.1$ K and in the presence of a magnetic field $h_{\mathrm{FC}}$.
The hysteresis loops have been computed cycling the magnetic field between $h= \pm h_{\mathrm{FC}}$ in steps $\delta h= 0.05$ K, and the magnetization components averaged at every field value during 100 MC steps, discarding the initial 100 MCS after every field step.  
\begin{figure}[t] 
\centering 
\includegraphics[width=0.5\textwidth]{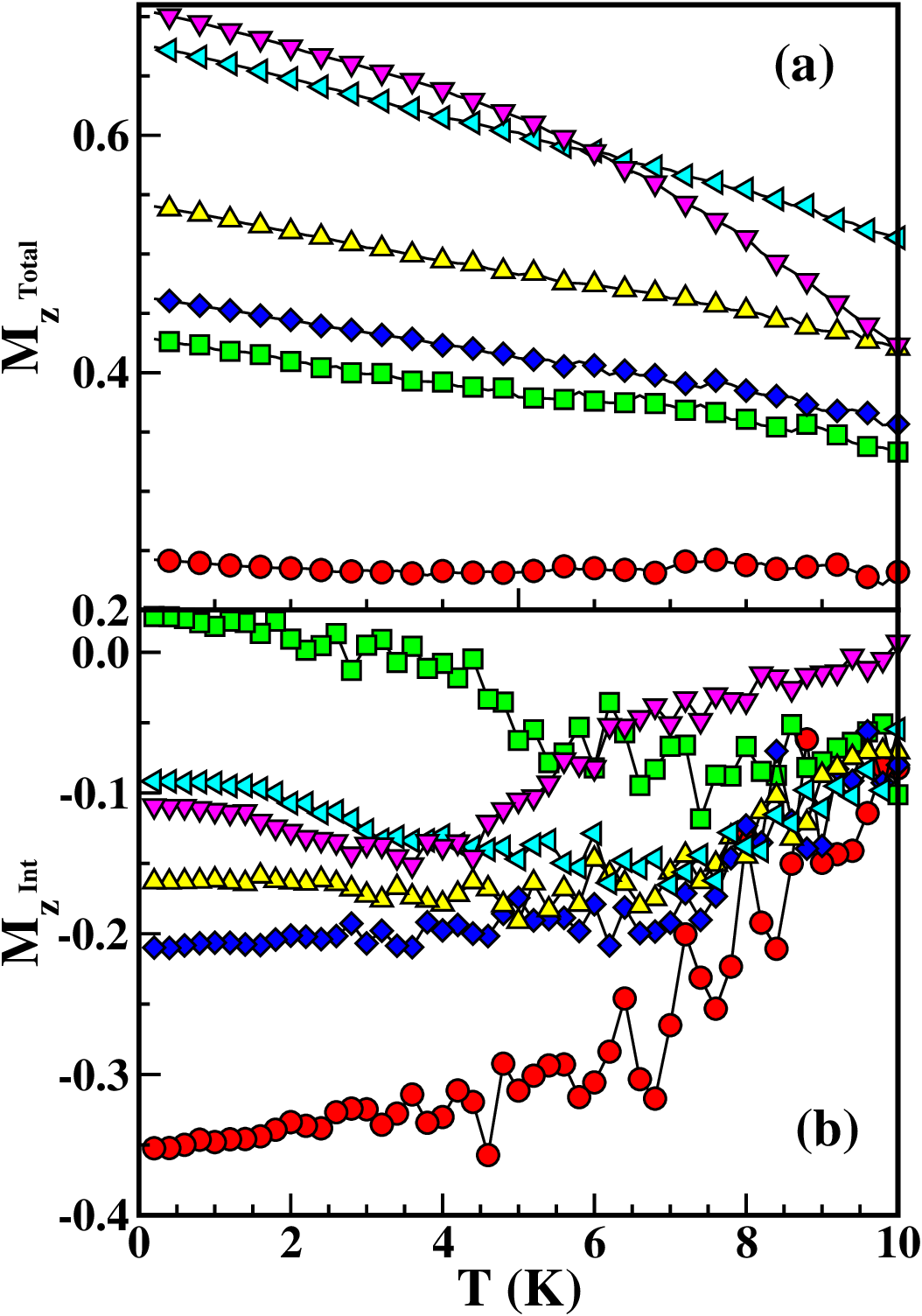}
\caption{(Colour online) Thermal dependence of the magnetization on cooling in the presence of a magnetic field applied along the z axis $h_{FC}= 4$ K for particles with $R_{Sh}= 3$a and different radii $R= 7, 9, 10, 12, 17, 20$a (from bottom-up). The value of the shell and core anisotropy constants are  $k_{\mathrm {S}}= 10$ and $k_{\mathrm{C}}= 1$, respectively. The values of the exchange couplings are: $J_{\mathrm C}= 10$ K, $J_{\mathrm S}= -0.5 J_{\mathrm C}$, and $J_{\mathrm {Int}}= -0.5 J_{\mathrm C}$. Panel (a) shows the total magnetization along the field direction, while panel (b) shows  the contribution corresponding only to the interfacial spins at the shell for the same selected radii (same symbols as in panel a).
}
\label{Fig1_fig}
\end{figure}

\section{Results}
\subsection{Particle size dependence}
The first objective of our work has been to study the dependence of the EB phenomenology on the particle size. In principle, the dependence of the exchange bias field $h_{\mathrm {EB}}$ should be similar to that observed in thin film systems on the thickness of the FM layer. If this  was the case, one should expect an increase of $h_{\mathrm {EB}}$ when reducing the particle size($h_{\mathrm {eb}}\sim 1/R_{\mathrm {C}}$). Some experiments have reported this trend in particle systems with different compositions \cite{Gangopadhyay_jap93,Peng_prb00,DelBianco_prb02}. However, other studies \cite{Vazquez_phab04,Mumtaz_jmmm07} reported the contrary variation in a certain range of particle sizes and some have even observed and argumented the disappearance of EB below a critical particle size \cite{Dobrynin_apl05,Boubeta_prb06}. 
In order to clarify the origin of these discrepancies, we have first studied the magnetic configuration attained when cooling nanoparticles with different total radii $R$ and the same shell thickness $R_{\mathrm {Sh}}= 3$ in the presence of a constant field. The results of the cooling process are displayed in the figure \ref{Fig1_fig}, where we show the thermal dependence of the total magnetization (panel a) and also the contribution of the interfacial shell spins $M_{\mathrm {Int}}^{\mathrm {Sh}}$ (panel b) for some representative particle sizes. 
The total magnetization increases with decreasing temperature, reflecting the progressive alignement of the FM core spins towards the field direction (see figure \ref{Fig1_fig}a). The values of magnetization attained at the lowest temperature increase with the particle size simply because of the increasing ratio of core spins with respect to the total number of spins in the particle.
In contrast, the low temperature net magnetization of the AFM shell spins at the interfacial region varies in a nonmonotonous way with the particle size (see figure \ref{Fig1_fig}b). Small changes in the particle core radius may induce very different geometric arrangement of spins at the interface due to the particular intersections of the spherical shape with the lattice sites. As a consequence, the magnetic configuration of interfacial shell spins after field cooling, induced below the Ne\'el temperatute of the AFM, is different for different particle sizes.

\begin{figure}[t] 
\centering 
\includegraphics[width=0.5\textwidth]{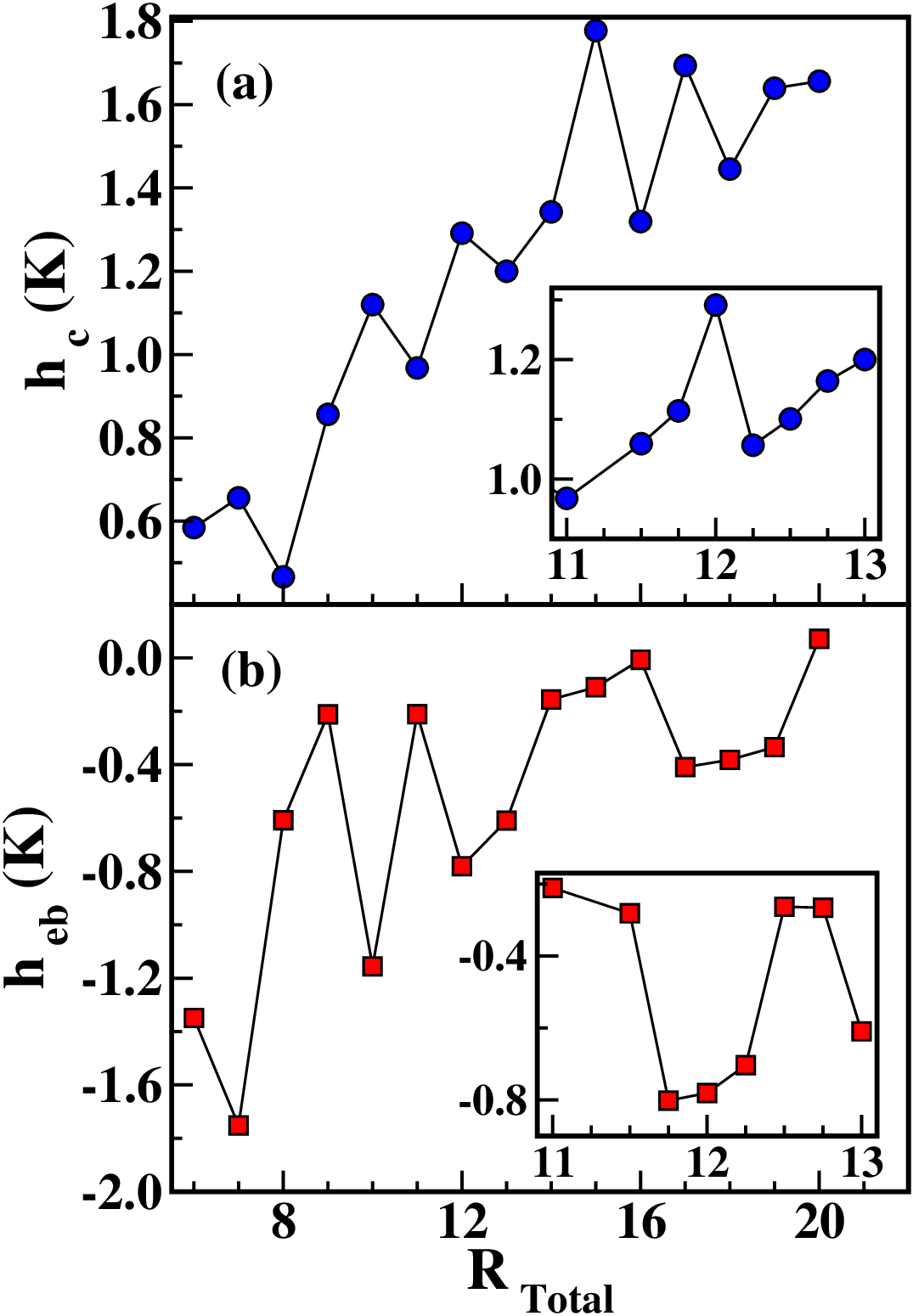}
\caption{(Colour online) Dependence of the coercive field $h_c$ (panel a) and exchange bias field $h_{\mathrm{eb}}$ (panel b) on the total particle radius $R_{\mathrm{Total}}$ in lattice parameter units) for a shell thickness $R_{\mathrm{Sh}}= 3$a. The insets show a detail of the main panels around the region $R_{\mathrm{Total}}= 11-13$a.
}
\label{Fig2_fig}
\end{figure}

In order to see the influence of these different magnetic states established after field coling for different particle sizes on the EB effect, we have also simulated hysteresis loops starting from the FC configurations for a range of particle size covering $R= 6- 20$a and computed the values of the coercive field $h_\mathrm{C}= (h_\mathrm{C}^+ - h_\mathrm{C}^-)/2$ and the shift of the hysteresis loops $h_\mathrm{eb}=(h_\mathrm{C}^+ + h_\mathrm{C}^-)/2$ from the values of the coercive field at the decreasing $h_\mathrm{C}^+$ and increasing field $h_\mathrm{C}^+$ branches. The dependence of these two quantities on particle radius are shown in figure \ref{Fig2_fig}. 
We observe that, when decreasing the particle size, there is a trend for decreasing $h_\mathrm{C}$ which can be attributed to the higher proportion of interfacial core spins that have to be reversed along the hysteresis loop with increasing particle size. 
This observation, is in agreement with our previous finding \cite{Iglesias_prb05} that the coercive fields after a FC process are higher than those obtained for zero field cooled hysteresis loops. 
For the EB field $h_\mathrm{eb}$, we observe that, although clear oscillations of this quantity with the particle size are seen, there is a trend to decrease as particle size increases. 
The oscillations are in clear correspondence to the ones observed in $M_{int}$ in figure \ref{Fig1_fig} and this is a demonstration that the microscopic origin of the loops shifts is related to the net magnetization component of the shell interfacial spins, since the local fields felt by the core spins are proportional to it, as we showed in our previous work \cite{Iglesias_prb05}. 
Therefore, depending on the particular lattice geometry and the configuration of nearest neighbours at the interfacial spins, very different values of $h_\mathrm{eb}$ can be obtained, even for particles having close values of the radius. In order to demonstrate this, we computed additional hysteresis loops for some selected particle radii between $11$a and $13$a. The corresponding results for $h_\mathrm{C}$ and $h_\mathrm{eb}$ are displayed in the insets of Fig. \ref{Fig2_fig}, where the above mentioned fluctuations in both quantities can be clearly seen even for very close values of the particle radii.
  
The reduction of the loop shift with particle size is in agreement with experimental observations reporting a similar trend \cite{Gangopadhyay_jap93,Peng_prb00,DelBianco_prb02}. However, the opposite trend reported in some cases \cite{Vazquez_phab04,Mumtaz_jmmm07} could be adscribed, within the scope of our model, to the fact that the limited range of particle sizes present in the samples of these experiments lain in between one of the oscillation periods found in our simulations giving rise to an increase in $h_\mathrm{eb}$ instead of the expected decrease. However, it is possible that the existence of interparticle interactions and distribution of anisotropy easy-axis in real samples could also influence the dependence of $h_\mathrm{eb}$ on the particle size.
\begin{figure}[tbp] 
\centering 
\includegraphics[width=0.5\textwidth,angle= 0]{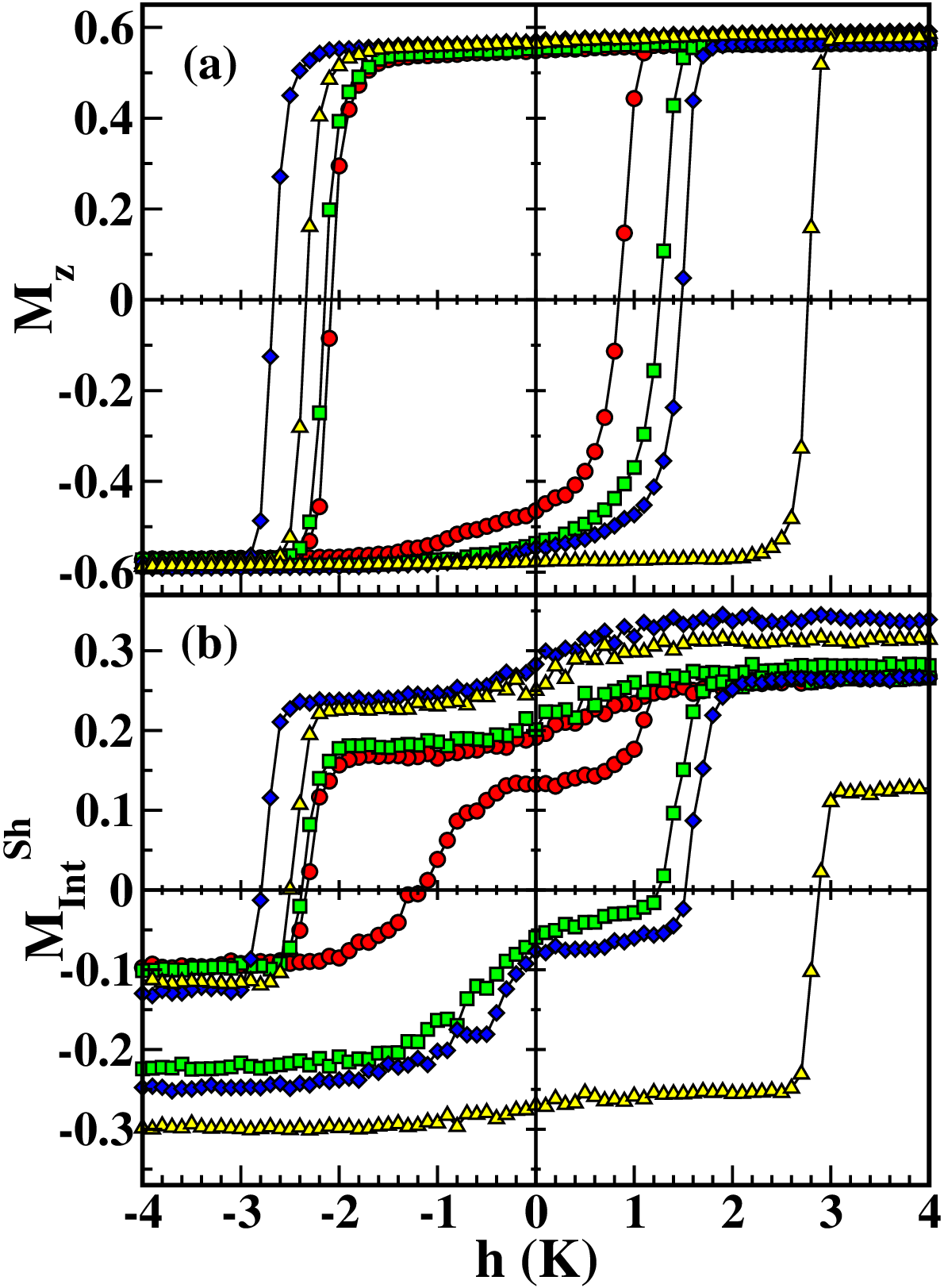}
\caption{(Colour online) Hysteresis loops for a particle of radius $R_{Total}= 12$ and ferromagnetic coupling at the interface of value $J_{Int}= +0.5 J_{C}$ obtained after field cooling from a high temperature in different magnetic fields $h_{FC}=$ 4 K (red circles), 15 K (green squares), 20 K (blue diamonds), 30 K (yellow triangles).
}
\label{Fig3_Fig}
\end{figure}
\subsection{Cooling field dependence}
It is well known from the experimental phenomenology on thin films formed by FM/AFM coupled bilayers, that the value and sign of the loop shifts may depend not only on the thickness of both layers but also on the nature (FM or AFM) of the coupling at the interface between them and on the value of the magnetic field applied during the FC process. For these kind of structures, it has even been demonstrated that loop shifts along the direction contrary to the cooling field (positive EB) can be obtained for high enough values of the cooling fields \cite{Nogues_prb00}.

Fewer studies about the $h_\mathrm{FC}$ dependence of EB have been reported for core/shell nanoparticle systems, and there is no clear-cut interpretation of the results. While a study on Co/CoO nanoparticles \cite{Zhou_apa05} found an increase of $h_\mathrm{eb}$ for $h_\mathrm{FC}$ values up to $5$ T, in another study of Fe/FeO nanoparticles \cite{DelBianco_prb04,Fiorani_jmmm06} the authors observed a non-monotonic dependence with a maximum $h_\mathrm{eb}$ at $h_\mathrm{FC} = 5$ T and a decrease above this value. 
%
\begin{figure}[tbp] 
\centering 
\includegraphics[width=0.5\textwidth,angle= 0]{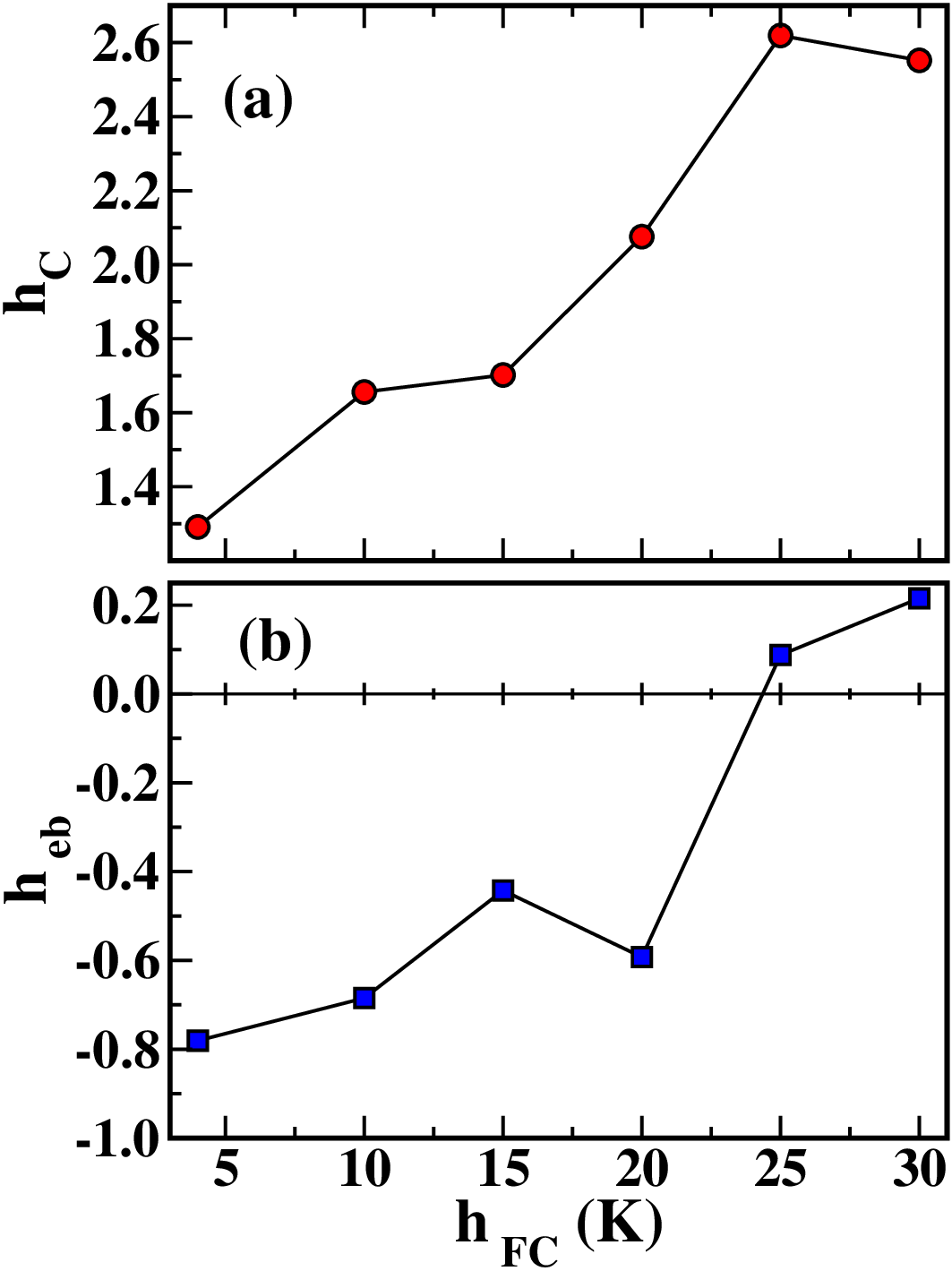}
\caption{(Colour online) Dependence of the coercive field $h_c$ (panel a) and exchange bias field $h_{eb}$ (panel b) on the magnetic field applied during the field cooling process previous to the calculation of the hysteresis loops for a particle with radius $R_{Total} =12$a.
}
\label{Fig4_Fig}
\end{figure}
In order to clarify these observations, we have performed simulations of hysteresis loops with different values of the cooling field $h_\mathrm{FC}$ for a particle of radius $R_\mathrm{Total}= 12$a and FM interfacial coupling $J_\mathrm {Int}= + 0.5 J_\mathrm {C}$. The results are presented in figure \ref{Fig3_Fig}a for some representative values of  $h_\mathrm{FC}= 4, 15, 20, 30$ K, only the central part of the loops is displayed. In all cases, the loops are closed for values of $|h|$ lower than $4$ K, but the shape of the loops becomes more symmetric with increasing $h_\mathrm{FC}$ as indicated by the increased squaredness and higher values of the decreasing field branch remanent magnetization. However, as it is already aparent by direct inspection, the shift of the loops (see figure \ref{Fig4_Fig}b, where the dependence of $h_\mathrm{eb}$ on $h_\mathrm{FC}$ is shown), which for low $h_\mathrm{FC}$ is in the negative direction, becomes positive for $h_\mathrm{FC}> 25$ K. In order to understand these observations, it is conveninent to look at the partiular contribution of the shell spins at the interface to the hysteresis loops, which are displayed in figure \ref{Fig3_Fig}b, that present a clear dependence on $h_\mathrm{FC}$.
First of all, let us notice that the magnetization values attained by the interfacial shell spins after the FC process previous to the start of the hysteresis loop ($M_\mathrm{Int}^\mathrm{Sh}$) increase with increasing $h_\mathrm{FC}$, as signaled by the increasing magnetization values of the descending loop branch and by direct comparison of the  $M_\mathrm{Int}^\mathrm{Sh}$ values at the end of the FC process. Therefore, the effect of an increasing $h_\mathrm{FC}$ is to reverse progressively the interfacial shell spins having one or two neighbours in the core into the field direction. The interfacial shell spins are the ones that generate the local exchange fields felt by the core spins. When cooling at low $h_\mathrm{FC}$, this exchange field points along a direction contrary to the magnetic field and produces a negative loop shift. Increasing $h_\mathrm{FC}$ values, reduce the value of the local exchange fields leading to a reduction of the loop shift and ultimately inducing exchange fields that point along the magnetic field direction and loop shifts into the positive direction (see figure \ref{Fig4_Fig}b).

It should also be noticed that the decreasing and increasing field branches of the loops for $M_\mathrm{Int}^\mathrm{Sh}$ in figure \ref{Fig3_Fig}b are quite different, with remanent magnetizations in the increasing branch lower (and even positive for the case $h_\mathrm{FC}= 4$ K) than for the decreasing one. This difference reflects the fact that a fraction of the shell interfacial spins remain pinned during the field reversal, a phenomenon that is at the microscopic origin of the EB effect \cite{Iglesias_prb05}. As the cooling field increases, the difference between the remanent magnetizations decreases in accordance with the reduction of $h_\mathrm{eb}$.  
Finally, the increase in coercivity with $h_\mathrm{FC}$ reported in figure \ref{Fig4_Fig}a, is caused by the higher fraction of spins at the interface that become aligned into the core magnetization direction and that have to be reversed by the magnetic field dragged by the core spins.  
The maximum in the $h_\mathrm{eb}$ vs $h_\mathrm{FC}$ curve reported in some works \cite{DelBianco_prb04,Fiorani_jmmm06} cannot be reproduced within the scope of our model. We think that, in order to explain this observation, additional features present in real samples, such as the policrystallinity of the AFM shell and surface disorder, should be incorporated into our model of nanoparticle, a work which is in progress at present. 
\section{Conclusion}

We have reported the results of simulations of the hysteretic properties of a model of a core/shell nanoparticle that have successfully explained the origin of some peculiar phenomenology associated to the EB effect. In particular, our results have linked the decrease of the loop shifts $h_\mathrm{eb}$ with particle size to the microscopic magnetic configurations established at the particle interface after field cooling, which are responsible also for the oscillatory behavior of $h_\mathrm{eb}$ with particle size found in the simulation results. We have also shown that, by varying the magnetic field applied during the FC process $h_\mathrm{eb}$, it is possible to modify the magnetic state of the spins at the interfacial region causing a corresponding decrease of $h_\mathrm{eb}$ and even loop shifts in the ositive field direction. We hope that further refinement of the model on which the simulations are based, by incorporating some source surface disorder will be able to account for training and time dependendent effects also observed experimentally.


\ack
We acknowledge CESCA and CEPBA under coordination of C$^4$ for computer facilities. This work has been supported by the Spanish MEyC through the MAT2006-03999, NAN2004-08805-CO4-01/02 and Consolider-Ingenio 2010 CSD2006-00012 projects, and the Generalitat de Catalunya through the 2005SGR00969 DURSI project. 


\end{document}